# A high stability semiconductor laser system for a $^{88}$Sr-based optical lattice clock

M. G. Tarallo, N. Poli, M. Schioppo, G. M. Tino*
*Dipartimento di Fisica and LENS–Università di Firenze,*
*INFN–Sezione di Firenze,*
*Via Sansone 1, 50019 Sesto Fiorentino, Italy*
(Dated: June 4, 2010)

We describe a frequency stabilized diode laser at 698 nm used for high resolution spectroscopy of the $^1S_0$-$^3P_0$ strontium clock transition. For the laser stabilization we use state-of-the-art symmetrically suspended optical cavities optimized for very low thermal noise at room temperature. Two-stage frequency stabilization to high finesse optical cavities results in measured laser frequency noise about a factor of three above the cavity thermal noise bewteen 2 Hz and 11 Hz. With this system, we demonstrate high resolution remote spectroscopy on the $^{88}$Sr clock transition by transferring the laser output over a phase-noise-compensated 200 m-long fiber link between two separated laboratories. Our dedicated fiber link ensures a transfer of the optical carrier with frequency stability of $7 \cdot 10^{-18}$ after 100 s integration time, which could enable the observation of the strontium clock transition with an atomic Q of $10^{14}$. Furthermore, with an eye towards the development of transportable optical clocks, we investigate how the complete laser system (laser+optics+cavity) can be influenced by environmental disturbances in terms of both short- and long-term frequency stability.

PACS numbers: 06.30.Ft, 42.62.Fi, 32.70.Jz, 37.10.Jk, 42.62.Eh, 42.55.Px

## I. INTRODUCTION

Optical frequency standards based on strontium atoms have demonstrated tremendous advantages in terms of short term stability (approaching $10^{-15}$ at 1 s [1–3]) and ultimate accuracy (approaching $10^{-17}$ level [4]) with respect to the microwave atomic counterparts. This improved level of precision enables new and more stringent tests of fundamental physics [5], and many new applications including improved timing for space travel and accelerator centers, as well as perhaps a new type of relativity-based geodesy [6]. Bosonic $^{88}$Sr-based optical clocks have demonstrated good features for compact and possibly transportable devices: easily accessible transitions for diode-based lasers for cooling and trapping the atoms, high natural abundance for high signal-to-noise ratio spectroscopy signals, low dead time for high duty cycles and a simplified spectroscopy scheme[7].

Presently, the optical local oscillator is the limiting factor for the Sr optical clock stability [2] via the Dick effect [8, 9]. Thus, improving the frequency pre-stabilization of these oscillators is a source of intense research in an effort to approach and surpass the $10^{-17}$ stability goal.

Rigid optical cavities are the most common devices used as short-term frequency references for optical standards. The employment of ultra-low thermal expansion materials as spacers between the cavity mirrors and the increasing level of engineering in their shape design have allowed several optical cavities to reach a stability level $\sigma_y \lesssim 6 \cdot 10^{-16}$ [10, 11] at 1 s. This instability level is limited by thermal fluctuations of the mirror substrates and coatings [12]. Although some theoretical ideas have been proposed to reduce or circumvent the thermal noise[13, 14], it currently sets the ultimate limit for cavity frequency stabilization techniques. Nevertheless environmental (acoustic, seismic, etc.) disturbances, poor spectral purity of the free running laser source, and thermal drifts can limit the optical local oscillator stability to much higher $\sigma_y$ values. Reaching the thermal-noise limit at the $10^{-16}$ level may require a complicated setup and considerable table space to isolate the reference cavity from environmental effects [15], which could be incompatible with the goal of transportable or space-based optical clocks [16]. These considerations have led us to develop a high-stability, compact clock laser for the Sr clock transition.

In this paper we describe in detail the clock laser source that we are developing as part of a compact optical frequency standard based on neutral strontium atoms [7]. Its main feature is its ultra-high finesse, vibration-insensitive design, and rigid optical cavity used as the short-term frequency reference. Our clock laser setup is one of the first to employ a rigid Fabry-Perot resonator with ULE spacer for thermal stability and high-reflectivity fused silica mirrors for low thermal noise, with a best expected instability value $\sigma_y = 3.8 \cdot 10^{-16}$ [17]. Other compact interrogation lasers have also been realized for strontium optical clock spectroscopy [18, 19].

This paper is organized as follows: in section II we present the setup of the clock laser system, in section III we describe the ultra-high-finesse cavities for stabilization of the clock laser at 698 nm, and in section IV we show and discuss the characterization measurements of the clock laser system. Finally, we show in the last section (V) the results of spectroscopy of $^1S_0$-$^3P_0$ clock transition with $^{88}$Sr and we present the achieved frequency instability of the active optical fiber link.

---

*Electronic address: Guglielmo.Tino@fi.infn.it

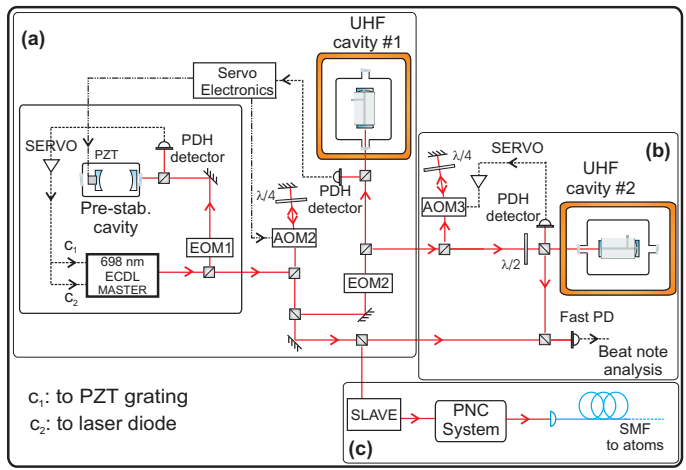

FIG. 1: Experimental setup for the 698 nm clock laser frequency stabilization and characterization. Block (a): master diode laser with two-stage frequency stabilization; (b): second independently-stabilized ultra-high finesse cavity and beat-note interferometer to make laser characterization studies; (c) noise-compensated fiber link which deliver interrogation laser to the atoms. UHF: ultra-high finesse; $AOM_i$: acousto-optic modulators; $EOM_i$: electro-optic modulators; SMF: single-mode optical fiber; PNC: phase noise cancellation system.

## II. FREQUENCY-STABILIZED DIODE LASER SYSTEM SETUP

The clock laser used to probe the $^1S_0$-$^3P_0$ strontium clock transition is a frequency stabilized diode laser at 698 nm. A schematic diagram of the system is shown in Fig.1. The whole system is housed in a clean room (class 10'000 ÷ 1'000) hundreds of meters away from the atomic reference laboratory. The clock laser is frequency-shifted and independently-locked to two ultra-high finesse cavities for frequency stabilization and system characterization, while an injection-locked slave diode amplifies the delivered power for optical frequency dissemination.

The optical system is partially insulated from seismic, acoustic and sub-acoustic disturbances. The optical table is supported by four pneumatic isolator legs. This system decouples floor seismic motions from the laser system, providing a -40 dB/decade vibration damping up to 30 Hz. A wooden panel above the table and a heavy-rubber curtain surrounding it block the direct clean room laminar air-flow and attenuate the acoustic noise from the rest of the laboratory.

The master laser is a commercial diode laser (Hitachi HL6738MG, non AR-coated) mounted in an extended cavity (Littrow configuration) with a length of 30 mm. The diode operates around 40 °C and delivers about 10 mW of optical power at 83 mA driving current. The free running ECDL exhibits a frequency noise spectral density $S_\nu(f) \simeq 4.6 \cdot 10^8/f$ (Hz$^2$/Hz) up to few MHz until it reaches its white noise plateau at $10^3$ (Hz$^2$/Hz) [20, 21]. Non-linear response and poor power coupling can occur when one attempts frequency stabilization on a high-finesse cavity. In particular, frequency noise below about 750 Hz yields large phase-modulation depths, while for frequencies higher than 20 kHz the frequency velocity becomes higher than the critical value for kHz-wide optical cavities[22]. We reduce the laser linewidth with two stages of Pound-Drever-Hall (PDH) frequency stabilization to optical cavities [23]. A one-step stabilization to an ultra-high finesse cavity is in principle possible by means of conventional servo electronics, but this would require 180 dB gain while keeping the same signal-to-noise ratio and the same amplifier fidelity level. Hence we use a first stabilization step to reduce the linewidth to < 1 kHz by locking the laser to a resonance of a pre-stabilization cavity. Then, we reduce further the linewidth by locking the pre-stabilized laser to a resonance of an ultra-high finesse cavity [15, 24].

The first pre-stabilization cavity is realized with an Invar spacer sitting on a v-shaped aluminum block. It has a finesse of $10^4$ and the resonance is about 150 kHz wide. The cavity length is tunable because one of the two end mirrors is attached to a piezoelectric transducer (PZT). In order to suppress externally-induced vibrations, the cavity is hermetically enclosed in a metallic pipe with anti-reflection (AR) coated wedges for optical access and a leaded-rubber layer to damp the transmitted vibration from the pipe. The servo correction signal is sent to two actuation channels. As shown in Fig. 1, one is sent to a PZT attached to the extended cavity grating, while the second goes directly to the diode current. The first loop has a single pole at zero and a gain of about 75 dB @ 10 Hz. The crossing frequency between the two loops occurs at about 100 Hz, with a combined servo bandwidth of about 2 MHz [21, 25].

For the second stage of frequency stabilization, the error signal generated by the PDH detector is sent to two parallel servo loops. One loop dominates at low frequencies (up to 1 kHz) and acts on the PZT of the pre-stabilization cavity, thereby compensating low frequency drifts. The second loop dominates at higher frequencies (up to 50 kHz) by acting on the double-pass $AOM_2$ (see Fig.1). The measured servo bandwidth of this second step of frequency stabilization is 54 kHz.

The frequency of the master laser can stay locked to the ultra-high finesse reference cavity for several hours as is needed for the remote spectroscopy of strontium atoms. Environmental thermal drifts limit the locking time by causing a pre-stabilization cavity drift that saturates the servo amplifier. A tight stabilization of the room temperature is achieved by two parallel temperature conditioning units. The room and the cavity tank temperatures are both recorded by a digital data logging system. For 24 hours data acquisition, the temperature distribution width of the temperature stabilization system in one day reaches almost the logger resolution (∼ 0.1 °C). The effect on the cavity error temperature shows a full-width half-maximum distribution close to the digitization resolution (∼ 1.6 mK). Further extension of the locking time could be achieved by controlling the temperature of the pre-stabilization cavity enclosure.

The entire stable-laser system (i.e., pre-stabilization cavity, ultra-high finesse cavity, master diode and optics) occupies an area of about $100 \times 80$ cm$^2$ which can easily be further reduced and acoustically isolated for transportable purposes. Nevertheless the results presented in this paper can be regarded as a full test for such applications.

## III. ULTRA-HIGH FINESSE CAVITIES FOR A HIGH STABILITY LASER

The key feature of our frequency-stabilized clock laser is the ultra-high finesse Fabry-Perot resonator used as a local frequency reference. The design of our two (identical) resonators meets three main requirements to minimize both the short- and long-term instability: high-reflectivity and very low mechanical loss mirrors, engineered geometry to reduce the impact of environmental vibrations, and low sensitivity to temperature fluctuations.

The ultra-high finesse cavity consists of a 10 cm long ULE (High grade Corning 7972 glass) spacer with two optically contacted fused silica mirrors. An ultra-narrow cavity resonance is ensured by the dielectric high-reflection coatings of the mirrors ($T_{mir} \sim 5$ ppm). The ULE spacer has a 'slotted' shape to allow the possibility to symmetrically support the cavity along the optical axis and reduce resonant frequency deviations due to mechanical vibrations.

The finesse of the cavity has been deduced both by measuring the photon cavity lifetime $\tau=43(2)$ $\mu$s and by directly observing the linewidth $\Delta\nu=3.7(0.5)$ kHz of the TEM$_{00}$ mode of the cavity. The two independent measurements give for both cavities a finesse of $4.1 \cdot 10^5$ within a 4% error. We also measured the on-resonance cavity transmittance $T_{FP} = 0.33$ which leads to $L_{mir} = 2.6 \pm 1.0$ ppm losses for one mirror and a cavity quality factor of $1.2 \cdot 10^{11}$, which represents the best result for an optical resonator at 698 nm to our knowledge [26]. This value also sets an upper limit on the mirror absorption and hence to the heating-induced frequency instability.

We have made an estimate of the thermal noise limit of our cavities based on the values reported for ULE, fused silica and mirror coatings in reference [12]. The resulting limit, $\sim 3.8 \cdot 10^{-16}$, is about three times smaller than the noise level in all-ULE vertical cavities realized at the same wavelength for a similar purpose [24].

Fig. 2 shows a schematic drawing of the cavity and its mechanical setup. The ultra-stable cavity sits on the aluminum-iron support inside a cylindrical vacuum chamber, which has maintains a pressure of $10^{-8}$ Torr with a 20 l/s ion pump. The cavity is supported horizontally with two aluminum arms connected by three low-expansion INVAR shafts. The effective supporting points are four square areas (about 2 mm$^2$ in size) with

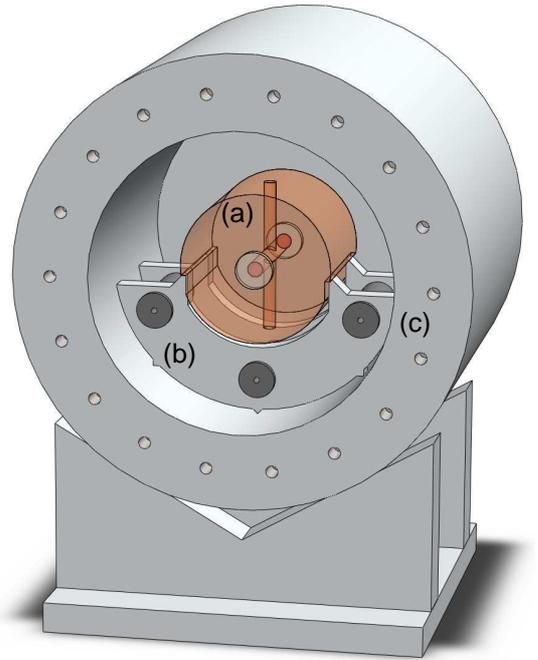

FIG. 2: Schematic drawing (to scale) of the ultra-high finesse cavity inside the vacuum system. The cavity (a) sits on four 2 mm$^2$ area on the aluminum support plate (b). To equilibrate the forces, small pieces of viton have been inserted between the cavity and the aluminum supporting areas. The aluminum plates are connected with three INVAR shafts and rest on the internal surface of the vacuum can on three points along three grooved lines (0.5 mm depth) that define the vertical direction inside the vacuum can (c). The thick aluminum vacuum chamber has an external diameter of 30 cm.

Viton square pieces (0.5 mm thickness) between the aluminum supporting points and the ULE spacer surface. The vacuum chamber is made of thick aluminum walls (5 cm) to increase the thermal inertia of the system.

The temperature of the outside surface of the vacuum can is actively stabilized at 25 °C by controlling the current passing through a high resistance (Alumel) wire that is wound around the can. Further insulation from environmental temperature gradients is provided by two-layers of polystyrene-and-plastic foil, which completely covers the tank. Finally the whole structure is contained in a 3 cm thick, extruded polystyrene box. This setup allows our heating system to keep a constant cavity temperature with a few tenths of a Watt flowing in the Alumel wire. The measured time constant of the system is about 3 hours. The amplitude of the residual control overshoot depends on the room temperature $T_{room}$ fluctuations. When $\Delta T_{room}$ is about 2–3 °C, the temperature variations of the external cavity tank are about 50 mK.

As with other state-of-the-art optical cavities used for laser frequency stabilization, our cavity geometry has been designed to reduce vibration sensitivity[27–29]. The ULE spacer has a "mushroom slotted shape" obtained by cutting an ULE cylinder along its longitudinal z-axis. The positions of the cuts along the horizontal and vertical direction are respectively $x = 46.19$ mm and $y = −6.08$ mm, where the axes origin is set on the center of the cylinder surface.

With the help of finite element method (FEM) simulations we calculated the cavity static distortion induced by accelerations in both vertical and horizontal directions, as a function of position of the supporting points along the longitudinal axis [17]. The resulting frequency shift is of the order of 2 kHz/(ms$^{-2}$) for acceleration along the vertical axis and 10 kHz/(ms$^{-2}$) for horizontal acceleration. For a typical vibration noise spectrum in a laboratory approaching 1 $\mu$g/$\sqrt{\text{Hz}}$, these values for the frequency sensitivity set a limit on the frequency noise of the order of 10 mHz/$\sqrt{\text{Hz}}$, more than a factor ten smaller than the calculated single cavity thermal noise $\sqrt{S_\nu^{TN}(f)} = 145/\sqrt{f}$ mHz/$\sqrt{\text{Hz}}$.

Good long-term frequency stability is due the small linear coefficient-of-thermal expansion (CTE) of the cavity, which result from the very low CTE's of the ULE spacer (0 ± 30 ppb/°C from 5°C to 35°C with a 95% confidence level[30]) and the fused silica mirrors (∼ 550 ppb/°C). We measured the cavity CTE near room temperature to be $7 \cdot 10^{-8}$/K and the length-change period to be about 3 hours. These values agree (within a factor 2 for the CTE) with a 2-dimensional FEM simulation of the cavity heat-transfer equation [31].

## IV. FREQUENCY STABILITY OF THE 698 NM DIODE LASER

In fig.3 the measurement of the frequency noise of the 698 nm clock laser source is shown. It has been measured by sending part of the light, frequency shifted by AOM$_3$ in Fig.1, to the second cavity, which rests on the same optical table, and analyzing the error signal obtained when the frequency of the beam is steered near the resonance of the second cavity.

The measured frequency noise reaches its minimum value of $\sqrt{S_\nu(f)} = 210$ mHz/$\sqrt{\text{Hz}}$ at 10 Hz and then exhibits a slow rise up to the servo bump at 50 kHz. At lower frequencies, the laser noise has a slope of $1/\sqrt{f}$ which is a factor of three higher than the thermal noise contribution of the cavity in the 2-11 Hz frequency band. This discrepancy maybe due to the influence of residual air currents in the beam path through the presence of parasitic etalons coupled to the high finesse cavity. This effect could be enhanced either by the quite large modulation frequency used for the PDH lock (14 MHz) and the long optical path between AOM$_3$ and the second high-finesse cavity (about 2 m)[10]. From the integration of this noise spectrum, we calculated a laser linewidth of about of 2 Hz.

To convert these noise measurements directly into frequency deviations and to check the residual relative cavity drifts, we also locked the frequency of the beam de-livered from the source to the second cavity using AOM$_3$ and generated a beatnote between the two beams. For convenience the resultant beat frequency has been down-converted from 200 MHz to about 3 kHz. The beatnote linewidth, recorded by a digital signal analyzer, is 3 Hz with a resolution bandwidth of 1.8 Hz. This value agrees with the spectrum in Fig.3. We measured the Allan deviation by counting the beatnote frequency with different gate times. The result (see Fig.4) is calculated by removing the linear drift with a computer controlled RF generator. The minimum $\sigma_y(\tau)$ is $1.1(3) \times 10^{-15}$ at $\tau = 67$ s.

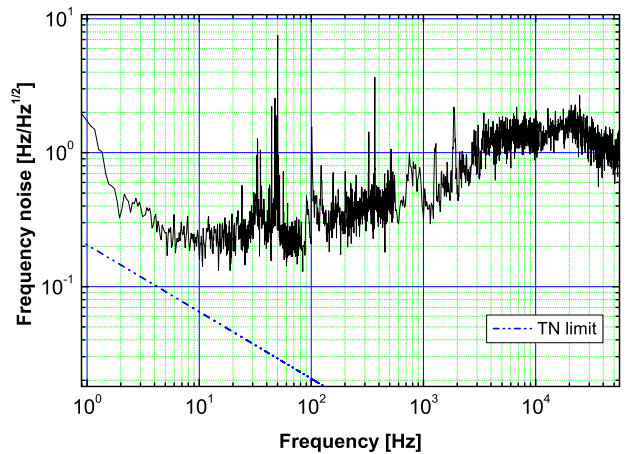

FIG. 3: Frequency noise of the stable 698 nm laser source locked to its ultra-high finesse cavity. The dashed line represents the calculated thermal noise (TN) limit due to the contribution of the ULE spacer, the fused silica mirror substrate and the Ta$_2$O$_5$/ SiO$_2$ coating. The spectrum takes into account the cavity pole.

We checked the the stabilized laser system's sensitivity to acceleration by observing the frequency noise imposed on the laser by accelerations in the vertical and horizontal direction. The results of the measurements are shown in Fig. 5. In the upper part of the figure the vibration noise spectra are included. The acceleration noise has been measured with a triaxial accelerometer (Kinemetrics Episensor), while the frequency noise has been measured using a resonance of the second ultra-high finesse cavity as a frequency discriminator. For comparison the frequency noise has been recorded with and without an air damping system applied to the optical table where the laser is mounted. With this floating system off, it is possible to recognize an excess of noise between 10 and 30 Hz, corresponding to the two main acceleration noise peaks in the horizontal and vertical directions. We looked for the peak values of the vibration noise which showed the highest spectral coherence between the single accelerometer channel signal and the frequency discriminator by means of a digital signal analyzer. We found the values of 3 kHz/(ms$^{-2}$) and 20 kHz/(ms$^{-2}$) for the sensitivity for vertical and horizontal directions, respectively. By using these coupling coefficients we can reconstruct the frequency noise due to environmental vibrations: as shown in Fig. 5, the frequency noise spectrum matches the reconstructed one between 7 and 20 Hz. The coupling coefficients are also in good agreement with the results of our FEM simulations.

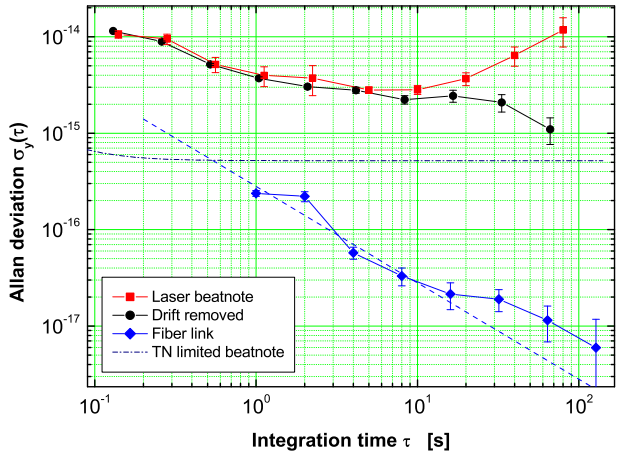

FIG. 4: Stability curves for the clock laser system. The plot shows the Allan deviation for the frequency-stabilized clock laser which approaches the thermal noise limit, while the fiber link does not limit the potential stability of the laser system.

A second possible contributor to our clock laser frequency instability is laser amplitude modulation noise. Due to the ultra-high finesse of the cavity, the typical optical power stored in the cavities is several watts. Slow laser power build-up and amplitude fluctuations can cause optical path length shifts via mirror thermal absorption[32]. Even assuming that the mirror losses are due only to absorption, at this power level photothermal shot noise is negligible [33]. Nevertheless, since the time constant of the thermoelastic effect for our cavity mirror geometry is of the order of 0.1 s (which corresponds to a cut-off frequency of 1.6 Hz) slow heating effects could dominate the nominal thermal drift.

We performed a measurement of the amplitude modulation noise sensitivity by locking the power of one of the cavities to a fixed voltage reference and looking for a frequency shift on the beatnote frequency when the reference value is changed. The final sensitivity measurement result is shown in Fig.6. The extracted slope determined the laser sensitivity to be

$$k_{AM} = 39.6 \pm 2.8 \text{ Hz}/\mu\text{W} \qquad (1)$$

This value can be compared with an estimate of the static thermoelastic effect, which depends on the mirrors' heating constant, its optical absorption and laser beam size. If we assume that all the mirror losses are due to absorption we get a static amplitude sensitivity $\delta\nu/\delta P \simeq 45$ Hz/$\mu$W [32], which is compatible with the experimental data. This value for $k_{AM}$ is large enough to require that the power of diode-based laser systems be actively controlled. For instance, for an input power of 300 $\mu$W, the relative intensity noise has to be at the level of $10^{-5}$ (-100 dB) to be equal to the TN noise limit at 1 s.

## V. REMOTE SPECTROSCOPY OF THE CLOCK TRANSITION

We have used our stable clock laser for remote spectroscopy of the $^{88}$Sr $^1S_0$-$^3P_0$ transition [7]. The experimental system is shown in Fig.7. In this section we describe the details of the coherent fiber link and the remote spectroscopy results.

### A. Low-noise optical frequency transfer by means of hundreds meter-long optical fiber

To deliver the 698 nm light from the stabilized laser to the Sr trap for spectroscopy on the $^1S_0$-$^3P_0$ clock transition we use a 200 m long, single mode optical fiber that connects two laboratories located in separated buildings.

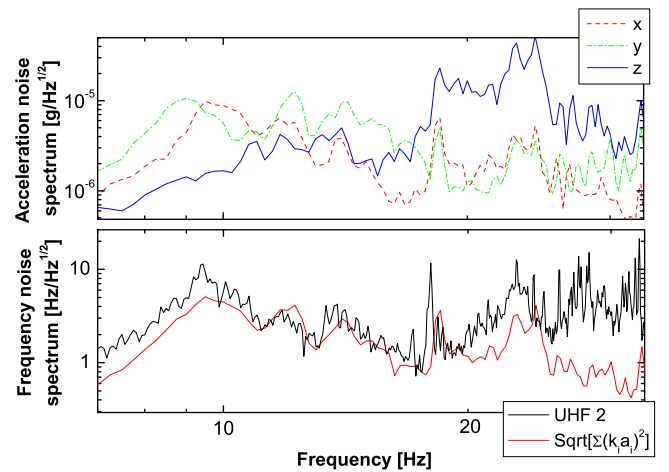

FIG. 5: Measurements of the clock laser sensitivity to accelerations. The upper plot shows the recorded acceleration noise spectra on the optical table without the air damping system (here z indicates the vertical with respect to the table). The lower plot shows the corresponding measured laser frequency noise (black line). Here the red line is the reconstructed frequency noise by using as sensitivity coefficients $k_y = 3$ kHz/(ms$^{-2}$) and $k_z$=20 kHz/ (ms$^{-2}$) respectively for vertical and horizontal directions.

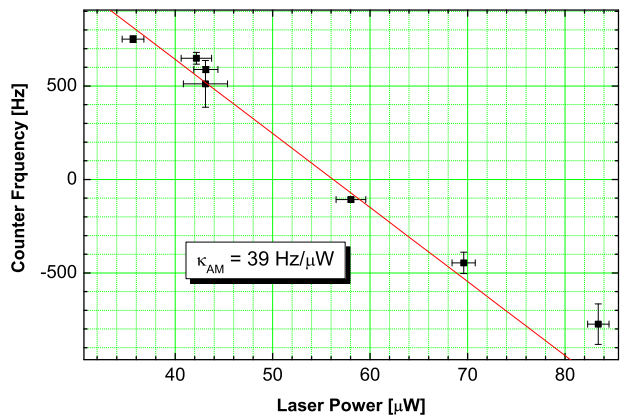

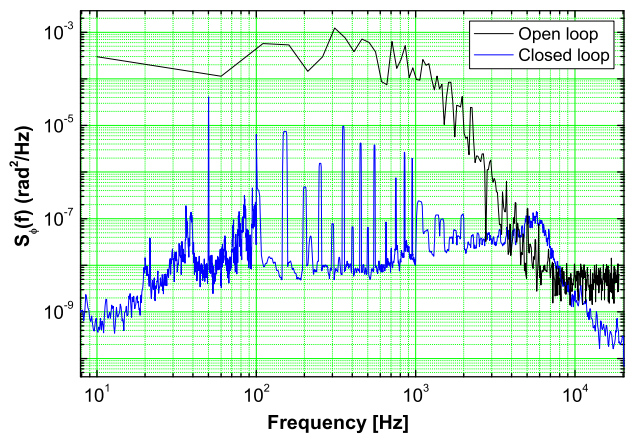

FIG. 6: Measurement of the frequency sensitivity to laser intensity shifts. The frequency counter measures the frequency displacement for different DC values of the servo-stabilized input power.

FIG. 8: Fiber phase noise cancellation servo loop performance. The plot shows the noise spectrum of the mixer output acquired by a digital signal analyzer. The open loop spectrum is obtained instead by converting the single sided spectrum of the beatnote from a spectrum analyzer.

The excess phase noise coming from the coupling of environmental noise to the fiber results in an effective broadening of the spectrum of the stabilized laser after one round trip of about 2 kHz. To preserve the spectral purity of the clock laser at the fiber end we then set up a fiber phase noise cancellation (PNC) system.

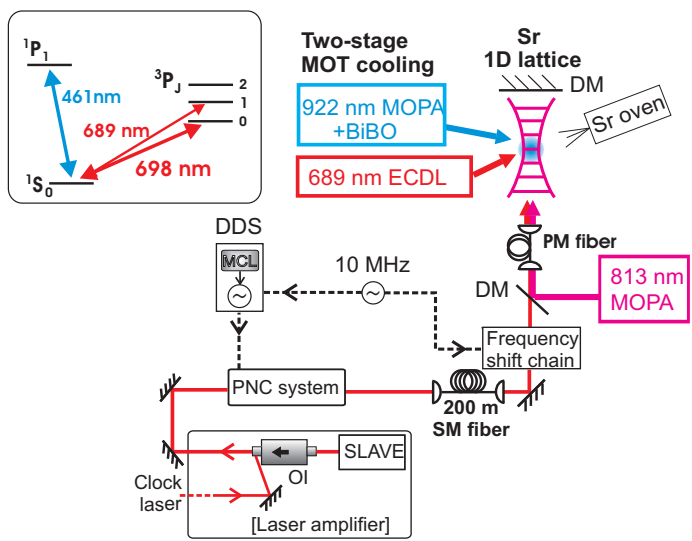

FIG. 7: Schematic diagram of the optical and electronic setup used to deliver and probe the clock transition of the lattice-confined $^{88}$Sr atoms. The stabilized laser is amplified by an injection-locked slave diode. The light is then split and frequency-shifted by an AOM inside the PNC for heterodyne beat note detection. Most of the light is sent to the fiber and coupled to the lattice light by means of a dichroic mirror (DM). The inset shows the relevant energy levels for $^{88}$Sr optical clock scheme. OI: optical isolator; MCL: programmable microcontroller.

About 1 mW of light coming from the stabilized master laser at 698 nm is amplified to a level of 20 mW with an optically injected slave laser. Following the scheme proposed in [34] this light is split into two beams by a polarizer beam splitter. Most of the light is sent to a single pass AOM that is used to shift the frequency of the master laser and cancel fiber-induced phase noise. The diffracted first order from the AOM is coupled into the 200 m long optical fiber. Due to a 50% fiber coupling efficiency and additional losses in the fiber we have about 5 mW exiting the other fiber end. A fraction of this light is reflected by the fiber end itself which is terminated with a standard FC/PC connection. The second beam is used as local oscillator for phase noise detection.

The interference signal is then observed on a photodiode on which the reference and the reflected beam are coupled together. With typical powers of 100 nW and 1 mW for the reflected power and the reference beam, respectively, we get a signal-to-noise ratio of about 30 dB with 10 kHz resolution bandwidth.

The error signal for the phase-locked loop is obtained by using a digital phase and frequency detector (PFD) and a microcontroller-controlled DDS as the reference oscillator. The error signal obtained is then fed back to a VCO that controls the AOM frequency. The servo bandwidth is naturally limited by the period of the light round-trip to a value of about 800 kHz. Actually, with about 7 kHz servo bandwidth we were able to cancel almost completely the phase noise added by the fiber link path. The effect of the noise cancellation is best illustrated by means of the power spectral density of phase fluctuations, $S_\varphi(f)$ [35] measured at the output of the PFD with a digital signal analyzer. The result is shown in Fig.8.

We studied the performance of the fiber link also in the time domain, by directly measuring the Allan deviation of the counted beat note frequency of the signal coming from the photodiode, (see Fig. 4). The Allan deviation of this signal converted into relative optical frequency units corresponds to $\sigma_y^{clock}(\tau) = 2.8 \cdot 10^{-16} \tau^{-1}$, which reflects the effect of the white noise-limited phase-lock. The fiber link has a final instability of $7 \cdot 10^{-18}$ after 100 s integration time.

### B. Probing of the $^{88}$Sr clock transition

The fiber link delivers a few milliwatts of frequency-stabilized, noise-compensated laser light suitable for probing the $^1S_0$-$^3P_0$ clock transition of bosonic $^{88}$Sr atoms. We use the technique of magnetic field-induced spectroscopy (MIS) to quench the forbidden transition [36], while the experimental setup we used to cool and probe the atoms has been described in Ref. [7].

Following the usual scheme for optical lattice clocks, about $4 \cdot 10^4$ cooled Sr atoms are trapped in an optical lattice, which allows Lamb-Dicke suppression of Doppler and recoil shifts, near the "magic wavelength" of 813.428 nm [37] to cancel AC Stark shift of the transition frequency. In Fig. 9 the highest resolution spectrum is shown. The atoms are probed by the clock light under a static magnetic field $B_0 = 1.3$ mT, a probe intensity $I = 20$ W/cm$^2$ and an interaction period $\tau = 200$ ms. The expected Rabi frequency is $\Omega = 2\pi \cdot 36$ Hz and a $\pi$ pulse duration of about 14 ms. Such overdriving of the clock transition yields a Lorentzian lineshape approximatively equal to

$$P_e(\Delta) = \frac{\Omega}{2} \frac{\Omega + \Gamma_l/2}{\Delta^2 + (\Omega + \Gamma_l/2)^2} \qquad (2)$$

where $\Delta$ is the frequency detuning from the atomic resonance, and $\Gamma_l$ is a loss coefficient that is added to the pure Rabi process [38], which can be due to laser coherence decay (finite laser linewidth and drifts) or collisional effects [39].

In order to minimize the effects due to residual frequency drifts in the clock laser, the cycle duration is set to 1 s, and we use a photomultiplier tube to collect the fluorescence from the ground-state atoms that are excited by resonant light at 461 nm. Further active cavity drift removal is applied by feeding a linear frequency ramp to the AOM (see Fig. 7) located inside the PNC system in front of the input end of the long fiber. The value of the ramp rate is set with a home-made front panel that controls the microcontroller frequency by checking the resonance peak drift on the oscilloscope. All the RF signals for driving the AOMs along the clock light path are locked to a 10 MHz reference signal that comes from a low phase-noise quartz oscillator (Oscilloquarz BVA) slaved to a rubidium standard, which for long term is steered by a GPS clock signal. Thus the residual laser drift is kept at the 0.1 Hz/s level.

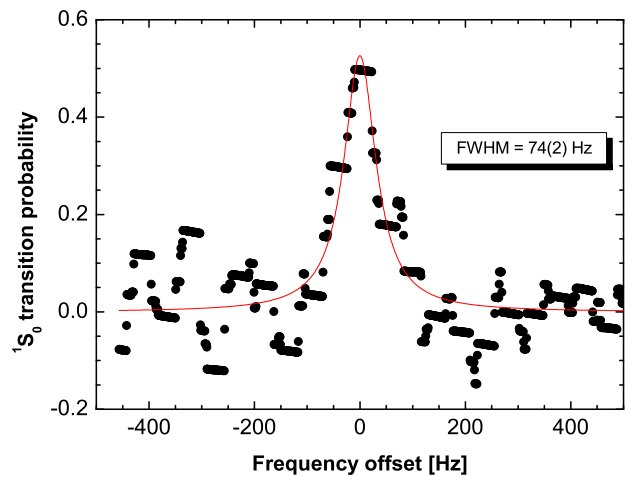

FIG. 9: High resolution spectrum of the $^1S_0$-$^3P_0$ transition taken by a single linear sweep of the probe laser frequency. The spectrum is obtained with a static magnetic field $B_0 = 1.3$ mT, a probe intensity $I = 20$ W/cm$^2$ and an interaction period $\tau = 200$ ms. The expected MIS Rabi frequency is $\Omega = 130$ Hz which limits the observed linewidth (see details on text). The signal-to-noise ratio is 7.

The result shown in Fig. 9 is compatible with a Rabi-limited linewidth: the discrepancy between predicted and experimental values is about 2 Hz, within the experimental error. This means that $\Gamma_l = (2 \pm 2)$ Hz. The peak density $n_0 = 2 \cdot 10^{11}$ atoms/cm$^3$ does not affect the spectral width of the clock transition [39], which implies that the measured linewidth is not collision-limited. This also means that the clock laser is compatible with an atomic quality factor $Q_{at} \leq 2 \cdot 10^{14}$, which should be attainable with minor changes in the present apparatus. The main limitation in the detection of the clock transition was due to fluctuations in the number of atoms trapped in the optical lattice. This can be suppressed by applying a normalization process that measures the population ratio of atoms in the lower/upper clock states by means of the $^3P_0$-$^3S_1$ transition at 679 nm to pump the excited atoms back to the $^1S_0$ state. A second improvement will be the amplitude and phase stabilization of the lattice laser.

### VI. CONCLUSIONS

We have presented the laser source we use for precision spectroscopy of $^1S_0$-$^3P_0$ intercombination line of strontium. We demonstrated that a simple, 0.8 m$^2$ optical system, can be characterized and stabilized at the Hz level for possible transportable optical clock applications. The measured frequency noise reaches its minimum value of 210 mHz/$\sqrt{\text{Hz}}$ at 10 Hz, a factor of three above the thermal noise limit. We addressed some specific problems such as effects of environmental vibration on the whole stabilization system, heating-induce noise on ultra-high finesse cavities employing fused silica mirrors, and remote transfer of optical frequency carrier over optical fibers. Remote MIS spectroscopy of the clock transition should not be limited by our clock laser at the Hz-level, allowing spectroscopy on the clock transition up to $Q_{at} = 10^{14}$.

## Acknowledgments

We thank G. Ferrari, R. Drullinger, F. Sorrentino, M. Bober for their work in the initial part of the experiment; A. Alberti for useful discussions; C. Oates and M. Prevedelli for their advices and careful reading of the manuscript; R. Ballerini, A. Hajeb, M. De Pas, M. Giuntini and A. Montori for technical assistance. This work was supported by LENS under contract RII3 CT 2003 506350, Ente CRF, ASI, and ESA.